%% file: main.tex
\documentclass[
 reprint, 
 superscriptaddress,
 nofootinbib,
 floatfix,
 amsmath,amssymb,
 aps,
 prl
]{revtex4-2}

\usepackage{hyperref}
\usepackage{graphicx}
\usepackage{xspace}
\usepackage{relsize}
\usepackage{comment}

\input babarsym

\begin{document}
\normalsize
\begin{flushleft}
\begin{minipage}{5cm}
\mbox{BABAR-PUB-26/02}\\
\mbox{SLAC-PUB-260812}\\
\end{minipage}
\end{flushleft}

\title{A New Precise Measurement of the \boldmath$e^+e^-\to\pi^+\pi^-(\gamma)$ Cross Section with the Initial-State Radiation Method at \babar}
\input{authors_feb2026_frozen_no_institutes_bad2022}


\begin{abstract}
We measure the $e^+e^-\to\pi^+\pi^-(\gamma)$ cross section from threshold to 1.4 GeV using 460 fb$^{-1}$ of \babar\ initial-state radiation data. The result is obtained from the $\pi^+\pi^-/\mu^+\mu^-$ ratio, reducing systematic uncertainties that are in common. Unlike the previous 2009 \babar\ analysis, the pion and muon contributions are separated statistically through their angular distributions. The compatibility with the previous \babar\ result is quantified including statistical and systematic correlations. The new measurement is also compared with other two-pion results as a function of energy and in terms of its hadronic vacuum polarization (HVP) contribution to the muon magnetic anomaly. The leading-order HVP contribution of the two-pion channel in the energy range of 0.3--1.4\gev is $a_{\mu}^{\pi\pi}=525.9\pm 3.5_\mathrm{stat}\pm3.8_\mathrm{syst}$ and $517.8\pm 2.4_\mathrm{stat}\pm3.3_\mathrm{syst}$ in units of $10^{-10}$ for the new 2026 and combined 2009 and 2026 \babar\ measurements, respectively.
\end{abstract}

\maketitle

\input{sections/introduction}\\
\input{sections/data_sample_method}\\
\input{sections/corrections_validation}\\
\input{sections/cross_section_hvp}\\
\input{sections/conclusion}

\input{babar_acknowledgements_reduced_mar2023.txt}

\bibliography{references}

\end{document}

%% file: authors_feb2026_frozen_no_institutes_bad2022.tex
\author{J.~P.~Lees}
\author{V.~Poireau}
\author{V.~Tisserand}
\author{E.~Grauges}
\author{A.~Palano}
\author{G.~Eigen}
\author{D.~N.~Brown}
\author{Yu.~G.~Kolomensky}
\author{M.~Fritsch}
\author{H.~Koch}\thanks{Deceased}
% ...
\collaboration{The \babar\ Collaboration}
\noaffiliation

%% file: sections/introduction.tex
\textit{Introduction}---
The leading-order hadronic vacuum polarization (HVP) contribution is one of the limiting ingredients in the Standard Model prediction of the muon magnetic anomaly from either data-driven evaluations or lattice QCD calculations~\cite{Aliberti:2025beg}. The data-driven prediction is obtained from dispersion integrals~\cite{Bouchiat:1961lbg} over measured $e^+e^-\to{\rm hadrons}$ cross sections, with the $e^+e^-\to\pi^+\pi^-$ channel providing the dominant contribution. The present experimental situation is unsettled: precise measurements in this channel are not mutually consistent, preventing progress through their straightforward combination~\cite{Aliberti:2025beg}. Independent measurements with different experimental strategies are essential for clarifying this picture and for reducing the uncertainty assigned to the hadronic contribution.

This Letter presents a new \babar\ measurement of the $e^+e^-\to\pi^+\pi^-(\gamma)$ cross section (including final-state radiation FSR) in the energy range from threshold to 1.4 GeV using initial-state radiation (ISR). In the ISR method, the emission of a hard photon from the incoming $e^+e^-$ state lowers the effective collision energy $\sqrt{s'}$ available to the hadronic or leptonic final system, allowing the full energy range relevant here to be measured with the same data sample collected near the $\Upsilon(4S)$ resonance. 

The analysis is based on the ratio of the pion and muon ISR spectra, $\pi^+\pi^-(\gamma)/\mu^+\mu^-(\gamma)$. 
The bare cross section for $e^+e^-\to\pi^+\pi^-(\gamma)$ is obtained from the pion spectrum and the effective ISR luminosity, 
\begin{equation}\label{eq:sigma0}
\sigma^0_{\pi\pi}(\sqrt{s'}) =
\frac{dN_{\pi\pi}/d\sqrt{s'}}
{\epsilon_{\pi\pi}(\sqrt{s'})\,dL^{\rm eff}_{\rm ISR}/d\sqrt{s'}} .
\end{equation}
 Through the same relationship between the muon spectrum and the QED bare cross section $\sigma^0_{\mu\mu}(\sqrt{s'})$, the $\mu^+\mu^-(\gamma)$ final state provides the effective ISR luminosity directly from data. 
 Several important effects cancel or are strongly reduced in the $\pi$-to-$\mu$ ratio, including the absolute $e^+e^-$ luminosity, the ISR-photon reconstruction efficiency, and vacuum polarization. The residual dependence on simulation is then concentrated in relative acceptances and data-driven efficiency corrections.

Radiative effects are especially important for ISR measurements. A recent \babar\ study of additional radiation found discrepancies with the prediction of the next-to-leading-order \textsc{Phokhara} Monte Carlo (MC) simulation~\cite{phokhara-v9.1} 
and demonstrated the importance of an inclusive treatment of extra photons~\cite{BaBar:2023xiy,Davier:2023fpl}.
The present measurement follows this inclusive strategy: events with additional ISR or FSR photons are retained whenever they are compatible with the signal topology. Since the dominant additional radiation is ISR, much of its impact cancels in the pion-to-muon ratio.

This measurement is complementary to the previous \babar\ result ~\cite{BaBar:2009wpw,BaBar:2012bdw}. The earlier analysis used particle identification to separate pions and muons and was based on approximately half of the full data set. Here the full 460 fb$^{-1}$ sample is used, and the pion and muon components are separated statistically through their distinct angular distributions in the two-particle center-of-mass (c.m.) frame. This avoids the dominant particle-identification limitations of the earlier measurement and allows a less restrictive track-momentum selection, providing a nearly independent and more inclusive determination of the cross section.

%% file: sections/data_sample_method.tex
\par\textit{Data Sample and Method}---
The analysis is based on the full \babar\ data set collected at the SLAC PEP-II asymmetric-energy $e^+e^-$ storage 
rings operated near the $\Upsilon(4S)$ resonance, corresponding to 424.2\,fb$^{-1}$ on resonance and 36.2\,fb$^{-1}$ 40\mev below~\cite{BaBar:2013agn}. The detector, described in Refs.~\cite{BaBar:2001yhh,BaBar:2013byz}, provides charged-particle tracking with the silicon vertex tracker and drift chamber inside a 1.5 T solenoidal field, photon detection in the CsI(Tl) electromagnetic calorimeter, and particle-identification information from several subsystems. In the present analysis, however, particle identification is not used to separate pions and muons in the signal extraction.

Candidate ISR events are required to contain a photon with energy $E_\gamma^\ast>4$\gev in the $e^+e^-$ c.m.\ frame and two oppositely charged tracks within the fiducial tracking acceptance, with a transverse-momentum threshold $p_\mathrm{T}>0.1$\gevc. In addition, the tracks are required to have at least 15 hits in the drift chamber, and to originate within $5\mm$ of the beam axis
and within $6\cm$ of the nominal collision point along the axis. 
Events may contain additional photons or additional tracks not satisfying the above signal track criteria; this maintains an inclusive selection with respect to extra radiation.

For each event, the charged-particle pair $X^+X^-$ is reconstructed under a common $m_X$ mass hypothesis. Kinematic fits to $e^+e^-\to X^+X^-\gamma{_{\rm{ISR}}\gamma}$ are used to improve the event reconstruction and to suppress multibody backgrounds. The kinematic fits account for either an additional photon detected at large angle or an undetected ISR photon emitted along the beam direction, and the event selection is based on the corresponding fit qualities. 

\begin{figure}[t]
  \centering
  \includegraphics[width=\columnwidth]{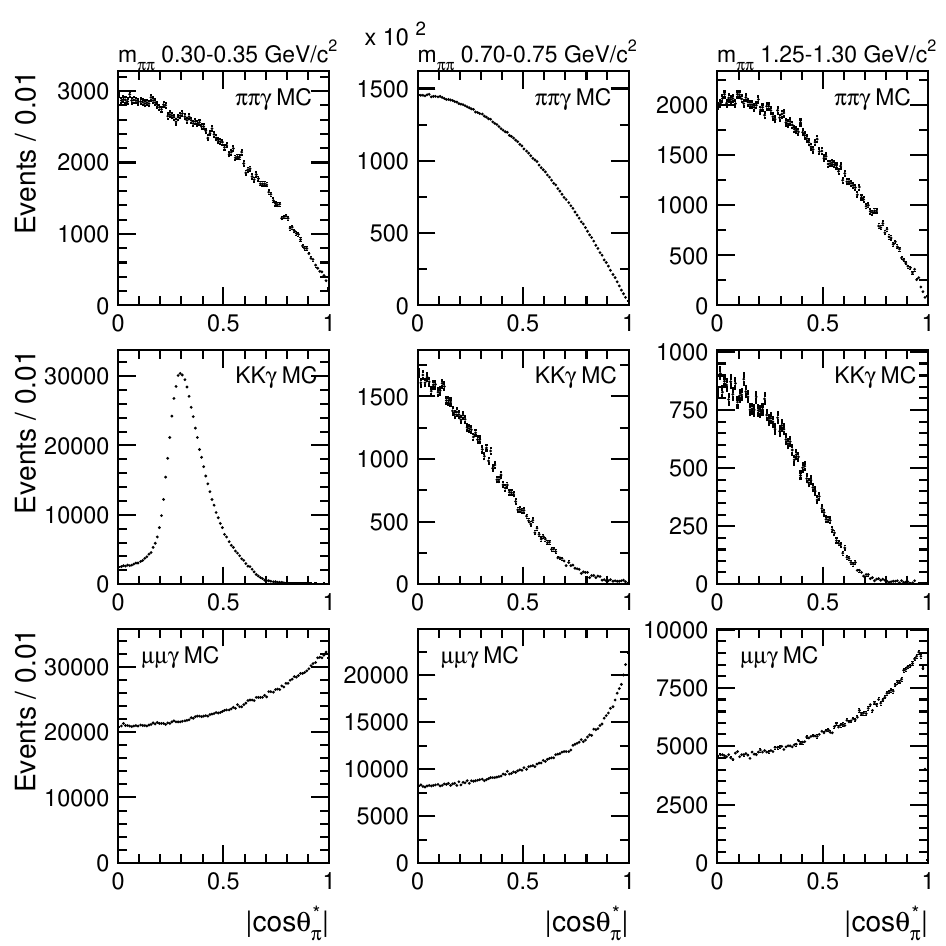}
  \caption{Simulated $|\cos\theta^\ast_\pi|$ distributions for
  $\pi^+\pi^-\gamma$, $K^+K^-\gamma$, and $\mu^+\mu^-\gamma$ events in
  three representative mass intervals within the detector acceptance. The charged
  tracks are reconstructed with the pion-mass hypothesis.}
  \label{fig:angular_templates}
\end{figure}

We define $\theta^\ast_X$ as the $X^-$ angle in the $X^+X^-$ rest frame with respect to the ISR-photon direction. The Born-level angular distributions are markedly different for spin-zero and spin-one-half particles: pion and kaon pairs follow approximately a $\sin^2\theta^\ast_{\pi,K}$ dependence, while muon pairs evolve from an almost flat distribution near threshold to the familiar $1+\cos^2\theta^\ast_\mu$ form at higher masses.\footnote{The quoted angular distributions can be found in elementary particle physics textbooks, e.g.,
Ref.~\cite{Halzen:1984mc}.} The large-angle ISR topology is particularly well suited to this method because the recoiling two-track system is boosted into the detector acceptance, preserving sensitivity over nearly the full angular range. 
The track momentum requirement, much less restrictive than the 1\gevc selection used in the earlier \babar\ measurement, is essential for retaining momentum-asymmetric two-body topologies, which populate the large-$|\cos\theta^\ast_X|$ region where the pion and muon angular distributions are most distinct. The resulting separation power in representative mass intervals is illustrated in Fig.~\ref{fig:angular_templates}.

After all selections, including the two-track mass interval restriction from threshold to 1.5\gevcc, the total data sample entering the angular fits amounts to 
7\,617\,066 events, of which 135\,249 events are estimated to be non-$\pi\pi\gamma$, -$\mu\mu\gamma$, or -$KK\gamma$ background.  
The pion and muon contributions are determined from template fits to the $|\cos\theta^\ast_\pi|$ distribution in bins of the reconstructed two-track invariant mass assuming a pion-mass hypothesis.
The kaon contribution, mainly distributed in the low mass region under the pion-mass hypothesis, is also included in the angular fits below 0.4\gevcc.
The $\pi^+\pi^-\gamma$, $\mu^+\mu^-\gamma$, and $K^+K^-\gamma$ templates are obtained from the \textsc{Phokhara} MC simulation 
and corrected 
for data to MC differences.
The templates of the residual $e^+e^-\gamma$ component, which receives contributions from radiative Bhabha events and photon conversions in regions that are not reliably simulated, are extracted from selected data control samples and the $e^+e^-\gamma$ contribution bin-by-bin normalization is obtained from the angular fits.

The angular fits use 10\mevcc\ mass intervals outside the $\rho$ region and 2\mevcc\ intervals between 0.5 and 1.0\gevcc.
The resulting two-pion mass spectrum of the $\pi^+\pi^-\gamma$ events is obtained directly as the distribution of pion yields determined in the angular fits. The dimuon mass spectrum, needed for the ISR-luminosity determination, 
is transformed from the pion-mass basis to the muon-mass basis using a 
transfer matrix derived from $\mu^+\mu^-\gamma$ simulation and validated with closure tests. 
A detailed description of the analysis can be found in~\cite{BaBar:2026prd}.

%% file: sections/corrections_validation.tex
\par\textit{Corrections, Unfolding, and Validation}---
The simulation is used to describe the geometrical acceptance, migration of events across mass bins, and the relative shapes of the angular templates. Residual differences between data and simulation are measured with control samples and propagated to both the  
template shapes and the derived mass spectra. The dominant corrections arise from the trigger, the charged-track reconstruction, the two-track vertex requirement, the kinematic-fit selection, and the momentum scale. Corrections depending on $|\cos\theta^\ast_\pi|$ are applied to the template shapes before fitting, while mass-dependent corrections are applied to the $\pi^+\pi^-\gamma$ and $\mu^+\mu^-\gamma$ mass spectra obtained from the angular fits.
Three independent sets of blinding factors are introduced for 
each pion and muon spectrum,
by applying blinded scale factors to tracking and trigger efficiencies measured in simulation (hence to the data absolute normalization), and to event yields obtained from the angular fits. 

Backgrounds not included as fitted two-body components are subtracted before the angular fits. Their normalizations are obtained from simulated samples corrected with measured cross sections and checked in data control regions. The dominant contributions are from ISR multihadron production, especially $\pi^+\pi^-\pi^0\gamma$ events, from continuum hadronic events at high mass, and from $p\bar p\gamma$ events close to threshold. Their combined uncertainty is small in the central mass region. 

As the cross-section measurement is obtained from the final-state spectra as a function of $\sqrt{s'}$ after ISR emission (Eq.~\ref{eq:sigma0}), the spectra derived from the angular fits are unfolded from reconstructed mass to $\sqrt{s'}$. 
Photons within $20^\circ$ of a charged track are classified as FSR~(\cite{BaBar:2023xiy}) and are included in the calculation of $\sqrt{s'}$.
The unfolding matrices relate the true final-state energy to the measured two-track invariant mass and describe the bin-to-bin migration of events due to detector smearing and FSR.  
They are built from simulation and used in an iterative unfolding procedure~\cite{Malaescu:2009dm,Malaescu:2011yg}. Closure tests with data-driven pseudo-experiments show no significant bias from the unfolding. 
Bootstrap replicates are used to propagate statistical uncertainties and their correlations through the full analysis procedure.\footnote{Several changes with respect to preliminary results of this analysis~\cite{Zhang:2026jvb,Polat:2026ysh} were incorporated 
for the final analysis presented here~\cite{BaBar:2026prd}.
The new results supersede the preliminary ones.}

The procedure 
provides an internal validation. 
The muon mass spectrum is compared with the QED expectation
as predicted by the \textsc{Phokhara} simulation of $\mu^+\mu^-(\gamma)\gamma_{\rm ISR}$ after applying the normalization factors and corrections that are needed for an absolute comparison,  
namely the ISR-photon reconstruction efficiency, the \babar\ luminosity, and the radiative correction to the next-to-leading order \textsc{Phokhara} generator. 
The analysis was unblinded only after verifying that the blinded data-to-QED ratio, $R_{\mu\mu}$, was compatible with a constant over the full mass range. The unblinded $R_{\mu\mu}$ and its constant fit are shown in the upper panel of Fig.~\ref{fig:pipi_results}, which gives
\begin{eqnarray}
    R_{\mu\mu}=0.9920&\pm&0.0030_{\rm stat}\pm0.0031_{\rm syst}\nonumber\\
    &\pm&0.0033_{\gamma{\rm ISR}}\pm0.0043_{{\rm lumi}\,ee},
\end{eqnarray}
with $\chi^2/\mathrm{ndf}=125/118$ (ndf standing for the number of degrees of freedom). This agreement validates the angular separation and correction procedure with a precision of 0.69\%. The dominant uncertainties in this absolute QED test, from the ISR-photon efficiency and the $e^+e^-$ \babar\ luminosity, cancel in the $\pi^+\pi^-/\mu^+\mu^-$ ratio used for the cross-section measurement.

%% file: sections/cross_section_hvp.tex
\par\textit{Pion Cross Section and HVP Contribution}---
The bare cross section for $e^+e^-\to\pi^+\pi^-(\gamma)$ is obtained according to Eq.~(\ref{eq:sigma0}) where the effective ISR luminosity is determined from the muon channel.
Equivalently, the measurement is driven by the unfolded $\pi^+\pi^-/\mu^+\mu^-$ ratio, multiplied by the QED prediction for the bare muon cross section and corrected for the pion acceptance. This construction preserves the main advantage of the ISR ratio method: the integrated $e^+e^-$ luminosity, the ISR-photon efficiency, and vacuum-polarization effects largely cancel between the two channels.
\begin{figure}[t]
  \centering
  \includegraphics[width=\columnwidth]{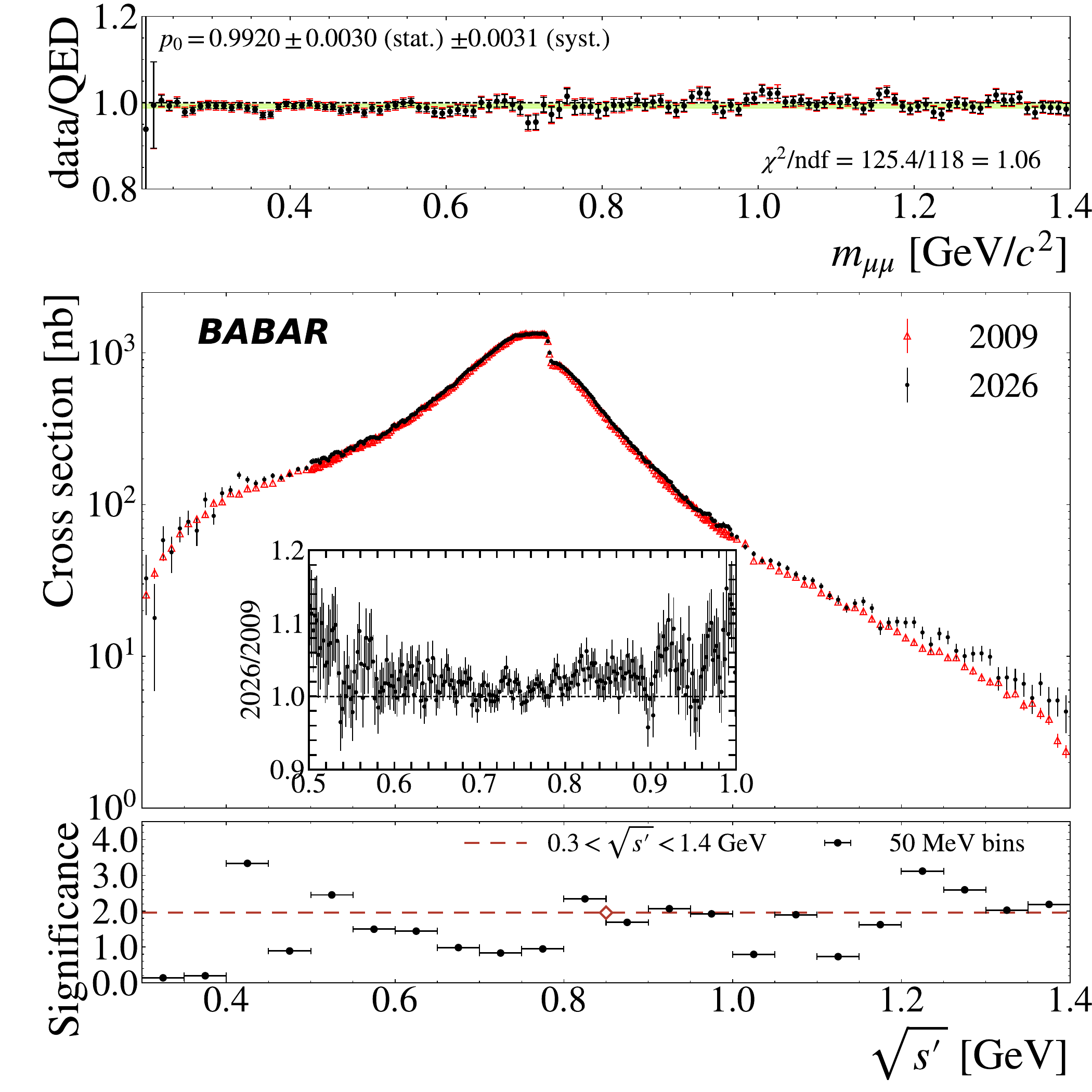}
  \caption{Upper panel:  
  Ratio $R_{\mu\mu}$ of the 
  $\mu^+\mu^-$ mass spectrum to the QED prediction.
  Middle panel: measured bare cross section for $e^+e^-\to\pi^+\pi^-(\gamma)$ from threshold to 1.4\gev in comparison with the previous 2009~\babar\ measurement~\cite{BaBar:2009wpw,BaBar:2012bdw}; inset: zoom of the central region (0.5--1.0\gev). Lower panel: local significance of the difference between the two \babar\ measurements, evaluated from 50 MeV-wide dispersion-integral intervals using the statistical and systematic uncertainties and their correlations across energy intervals.}
  \label{fig:pipi_results}
\end{figure}

The measured cross section is shown in the middle panel of Fig.~\ref{fig:pipi_results}. The statistical and systematic covariance matrices include the uncertainties from the angular fits, efficiency corrections, background subtraction, unfolding, and ISR-luminosity determination, with correlations propagated through the full analysis chain. In the central $\rho$ region, where the dispersive kernel and the large cross section make the measurement most important for HVP, the precision is dominated by the statistical separation of the pion and muon components and by the data-driven corrections to the template shapes. The uncertainties are energy dependent with the dominant systematic uncertainty sources all below 0.4\% in the $\rho$-peak region. Figure~\ref{fig:pipi_results} also compares the result with the previous \babar\ measurement.
The present analysis is statistically and systematically complementary to the previous study, as explained above. 
The new cross section is about 2\% higher than the 2009 \babar\ result in the $\rho$-resonance region, with larger fluctuations in the threshold and high-mass regions where the angular separation has lower statistical power and suffers from the overwhelming dominance of the muon contribution. The bottom panel of Fig.~\ref{fig:pipi_results} shows a scan of the local significance of the difference with the 2009 measurement, computed from 50\,MeV-wide dispersion-integral intervals using the uncertainties and their correlations across energy intervals. Treating the two measurements as independent, the differences are typically below 2 standard deviations in the central region. Bootstrap studies indicate small statistical correlations between the two measurements, while conservative systematic-correlation scenarios change the local significances by at most 0.2 standard deviations.

The contribution to the muon magnetic anomaly from the two-pion channel $a_{\mu}^{\pi\pi}$ is computed by integrating the measured cross section with the standard dispersive kernel~\cite{Brodsky:1967sr}. Table~\ref{tab:amu_pipi} gives the contributions in three energy intervals and compares the new results with the previous \babar\ ones and their combination, taking into account their correlation and applying uncertainty scaling when the local $\chi^2$ value is greater than unity~\cite{BaBar:2026prd}. The full 0.3--1.4\gev interval has statistical and systematic precisions of 0.67\% and 0.72\%, respectively. In the dominant 0.5--1.0\gev region, the corresponding precisions are 0.57\% and 0.56\%. However, in the $\rho$ tail regions below 0.5\gev and above 1\gev, where the $\pi\pi$ signal is overwhelmed by the large $\mu\mu$ cross section, the new measurement is not competitive with the 2009 result.
\begin{table}[tb]
\centering
\caption{Contributions to $a_\mu^{\pi\pi}$ per energy ranges, in units of $10^{-10}$. The first and second uncertainties are statistical and systematic, respectively.}
\label{tab:amu_pipi}
\resizebox{\columnwidth}{!}{%
\begin{tabular}{cccc}
\hline
Energy range  & Present  & BABAR & Combination \\
 (GeV) & result & 2009 & 2009+2026 \\
\hline
0.3--0.5 & $60.7(1.7)(1.4)$ & $57.6(0.63)(0.55)$ & $58.2(0.86)(0.78)$ \\
0.5--1.0 & $454.6(2.6)(2.5)$ & $445.9(2.1)(2.5)$ & $449.7(2.1)(2.5)$ \\
1.0--1.4 & $10.6(0.16)(0.34)$ & $9.7(0.10)(0.08)$ & $9.9(0.12)(0.14)$ \\\hline
Total & $525.9(3.5)(3.8)$ & $513.3(2.2)(3.1)$ & $517.8(2.4)(3.3)$ \\
\hline
\end{tabular}}
\end{table}

\begin{figure}
    \centering
    \includegraphics[width=\linewidth]{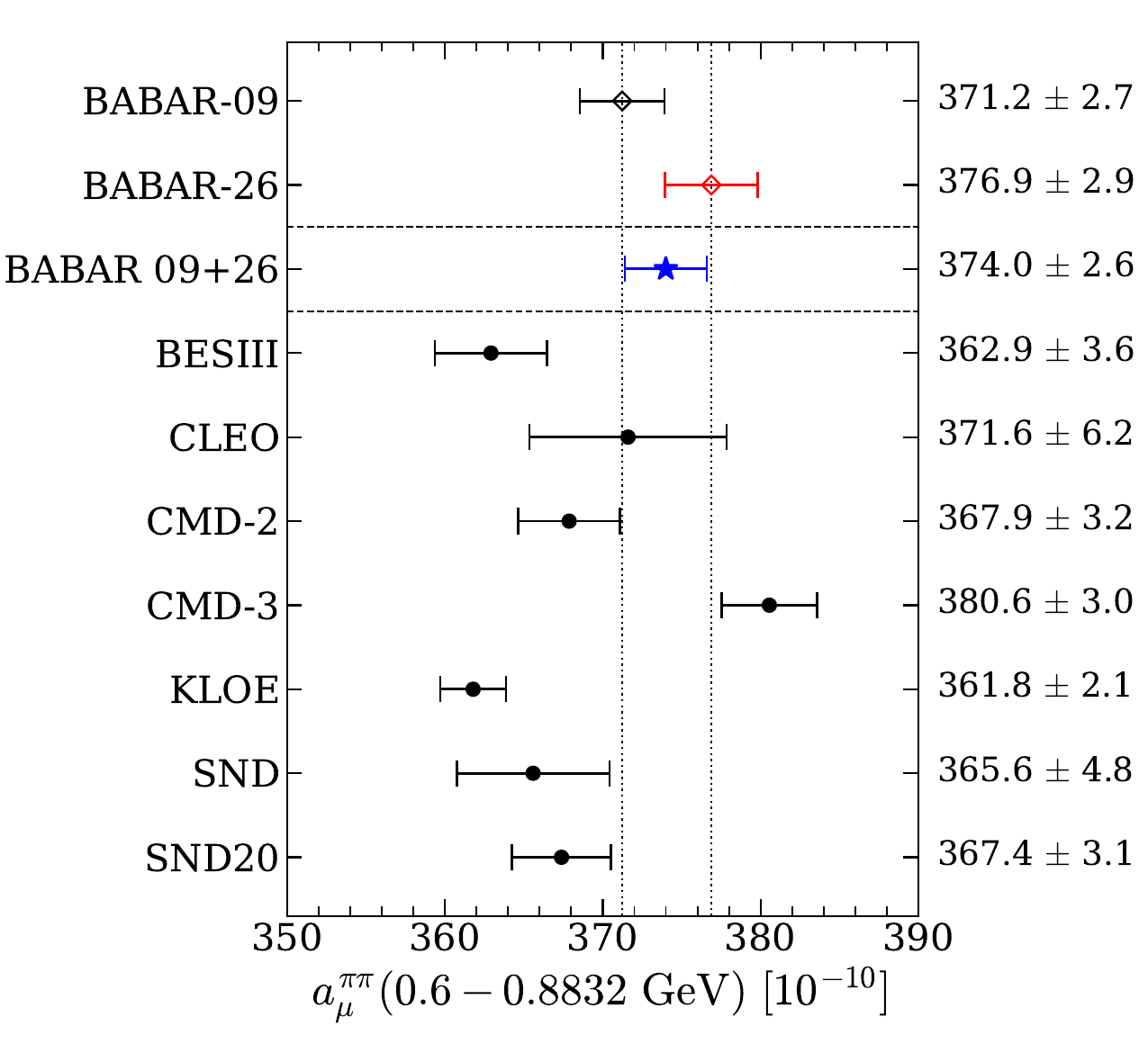}
    \caption{Comparison of the leading-order hadronic vacuum polarization contribution, $a_{\mu}^{\pi\pi}$, from $e^{+}e^{-}\!\to\!\pi^{+}\pi^{-}$ data in the 0.6--0.8832\gev energy range. Error bars denote the quoted experimental uncertainties. The two dotted vertical lines along the 2009 and 2026 \babar\ results are drawn to facilitate the comparison with the other determinations}
    \label{fig:amu_integrals}
\end{figure}

The new \babar\ result and the 2009+2026 combination are directly compared to the other recent determinations~\cite{BaBar:2009wpw,BaBar:2012bdw,BESIII:2015equ,Xiao:2017dqv,CMD-2:2003gqi,CMD-2:2005mvb,Aulchenko:2006dxz,CMD-2:2006gxt,CMD-3:2023alj,KLOE:2008fmq,KLOE:2010qei,KLOE:2012anl,Achasov:2006vp,SND:2020nwa} of $a_{\mu}^{\pi\pi}$ in the common region of measurements between 0.6 and 0.8832\gev in Fig.~\ref{fig:amu_integrals}. 
This 2026 result is within two standard deviations of most previous results, including the 2009 \babar\ ($-1.4\sigma$) and the most recent CMD-3 ($+0.9\sigma)$ results, but is inconsistent with KLOE ($-4.2\sigma$) and BESIII ($-3.0\sigma$). It is consistent with $\tau$-based estimates~\cite{Davier:2023fpl,Castro:2024prg} and recent lattice calculations~\cite{Aliberti:2025beg,Boccaletti:2024guq}.

%% file: sections/conclusion.tex
\par\textit{Conclusion}---
Using the full 460 fb$^{-1}$ \babar\ data sample collected on and below the $\Upsilon(4S)$ resonance, we have measured the $e^+e^-\to\pi^+\pi^-(\gamma)$ cross section from threshold to 1.4\gev with the ISR method. The analysis separates the $\pi^+\pi^-\gamma$ and $\mu^+\mu^-\gamma$ final states statistically through their angular distributions rather than by particle identification, allowing a less restrictive low-momentum track selection and providing a measurement essentially independent of the previous \babar\ result.

The $\mu^+\mu^-(\gamma)\gamma_{\rm ISR}$ spectrum validates the method through a complete comparison with QED, giving $R_{\mu\mu}=0.9920$ with a total precision of 0.69\%. The $e^+e^-\to\pi^+\pi^-(\gamma)$ cross section is then obtained from the unfolded $\pi^+\pi^-/\mu^+\mu^-$ ratio, preserving the cancellation of the ISR-photon efficiency, integrated luminosity, and vacuum-polarization uncertainties. This new measurement is a complementary \babar\ input to the data-driven HVP determination. The resulting two-pion contribution to the muon magnetic anomaly amounts to $525.9(3.5)(3.8)\times 10^{-10}$ in the 0.3--1.4\gev range where the first and second uncertainties are statistical and systematic, respectively. When combined with the \babar\ 2009 measurement in the same energy range, the corresponding result is $517.8(2.4)(3.3)\times 10^{-10}$.

%% file: babar_acknowledgements_reduced_mar2023.txt
\par\textit{Acknowledgements}---
We are grateful for the extraordinary contributions of our PEP-II colleagues in achieving the excellent luminosity and machine conditions that have made this work possible. The success of this project also relies critically on the expertise and dedication of the computing organizations that support \babar, including GridKa, UVic HEP-RC, CC-IN2P3, and CERN. The collaborating institutions wish to thank SLAC for its support and the kind hospitality extended to them. We also wish to acknowledge the important contributions of J.~Dorfan and our deceased colleagues E.~Gabathuler, W.~Innes, D.W.G.S.~Leith, A.~Onuchin, G.~Piredda, and R. F.~Schwitters. This work is supported by ANR-22-CE31-0011.